%% file: main.tex
\documentclass[%
 reprint,
 nofootinbib,
 amsmath,amssymb,
 aps,
 superscriptaddress,
]{revtex4-2}
\UseRawInputEncoding
\usepackage{graphicx}
\usepackage{float}
\usepackage{url}
\usepackage{dcolumn}
\usepackage{bm}

\usepackage{lipsum}
\usepackage{color}
\usepackage{hyperref}

\newcommand {\s}[1]{$\sqrt{s} = #1$}
\newcommand {\pp}{$p$+$p$}
\newcommand {\pt}{p_{\rm T}}
\newcommand {\nse}{n\sigma_{\rm e}}
\newcommand {\effPho}{\varepsilon_{\rm PHE}}
\newcommand {\nPho}{N_{\rm PHE}}

\newcommand {\jpsi}{J/\psi}

\newcommand {\Ke}	{K_{\rm e3}}

\begin{document}

\preprint{APS/123-QED}

\title{Measurement of inclusive electrons from open heavy-flavor hadron decays in $p$+$p$ collisions at \s{200} GeV with the STAR detector}

\input{star-author-list-2021-07-14.aps.tex}

\date{\today}

\begin{abstract}
We report a new measurement of the production cross section for inclusive electrons from open heavy-flavor hadron decays as a function of transverse momentum ($\pt$) at mid-rapidity ($|y|<$ 0.7) in $p$+$p$ collisions at \s{200} GeV. The result is presented for 2.5 $<\pt<$ 10 GeV/$c$ with an improved precision above 6 GeV/$c$ with respect to the previous measurements, providing more constraints on perturbative QCD calculations. Moreover, this measurement also provides a high-precision reference for measurements of nuclear modification factors for inclusive electrons from open-charm and -bottom hadron decays in heavy-ion collisions.
\end{abstract}

\maketitle


\section{\label{sec:introduction}Introduction}

In $p$+$p$ collisions, heavy (charm and bottom) quarks are dominantly produced in initial hard parton scatterings. As the masses of the heavy quarks ($m_{\rm Q}$) are much larger than the QCD scale parameter $\Lambda_{\rm QCD}$, such partonic scattering processes can be calculated by perturbative QCD (pQCD) down to low transverse momentum ($\pt$). Detailed study of heavy-quark production and the comparison to experimental data allows us to test pQCD in different kinematic regions ($\pt < m_{\rm Q}$, $\pt \sim m_{\rm Q}$, $\pt >> m_{\rm Q}$). It also provides an important testing ground for the pQCD calculations that deal with other processes of high interest involving multiple hard scales ($m_{\rm Q}$, $\pt$) such as weak boson production, Higgs boson production and physics beyond the Standard Model~\cite{ref:qcd,ref:QCDCB,ref:qcd1}. 

In heavy-ion collisions, heavy quarks are produced before the creation of Quark Gluon Plasma (QGP)~\cite{ref:QGP1, ref:QGP2} because their masses are much larger than the temperature of QGP, $\rm T_{\rm QGP}$. Heavy quarks experience all stages of QGP evolution and thus are an excellent probe of the QGP~\cite{ref:hf}. Heavy quark production in $p$+$p$ collisions serves as an important reference to compare with that in heavy-ion collisions for understanding the nature of interactions of heavy quarks with the QGP and the parton energy loss mechanisms in general. Significant suppression of the charm meson yield at large $\pt$, resulting from the substantial energy loss of heavy quarks in the QGP, has been observed at both RHIC and the LHC~\cite{hfreview1, hfreview2, D0_STAR, D0_ALICE, D0_ALICE1, D0_CMS}, indicating significant interactions between heavy quarks and the medium.  

Charmed hadron production in $p$+$p$ collisions has been measured by the Solenoidal Tracker at RHIC (STAR)~\cite{STAR:D0}, and was found to be consistent with the upper limit of Fixed-Order Next-to-Leading Logarithm (FONLL) calculations~\cite{ref:QCDCB}. Due to a large combinatorial background, these measurements have a limited $\pt$ range ($\pt <$ 6 GeV/$c$). Electrons\footnote{Unless specified otherwise, electrons referred here include both electrons and positrons and results are presented as $\frac{e^++e^-}{2}$.} from semileptonic decays of heavy-flavor hadrons, referred to as Heavy Flavor Electrons ($\rm HFEs$), have also been used as proxies to measure heavy quarks~\cite{STAR:NPE,NPE_ALICE,ref:OHFP}. Although kinematic information regarding the parent heavy-flavor hadrons is incomplete, and the $\rm HFE$ sample is usually a mixture of electrons from both charm and beauty hadron decays, $\rm HFE$s are still widely used to study heavy quark production because they have higher branching ratios\footnote{$D^0 \rightarrow e^++X$ (BR = 6.5\%), $D^+ \rightarrow e^++X$ (BR = 16.1\%), $B^0 \rightarrow e^++X$ (BR = 10.1\%), $B^+ \rightarrow e^++X$ (BR = 10.8\%).} than the hadronic decays of open heavy-flavor hadrons and data collection of high-$\pt$ electrons can be enhanced in an experiment by dedicated triggers. The inclusive $\rm HFE$ production in $p$+$p$ collisions at \s{200} GeV has been studied by the STAR~\cite{STAR:NPE} and PHENIX~\cite{PHENIX:NPE} experiments at RHIC. These earlier results are seen to be consistent with pQCD FONLL calculations, however their constraining power at high $\pt$ is limited by the large experimental uncertainties.

In this paper, we report a new measurement of the inclusive $\rm HFE$ cross section at mid-rapidity ($|y|<$ 0.7) in $p$+$p$ collisions at \s{200} GeV. The cross section for inclusive $\rm HFE$ as a function of $\pt$ ($2.5 <\pt< 10\, \rm GeV/c$) is obtained, with a higher precision at $\pt > 6\,\rm GeV/c$ than the previously published results~\cite{STAR:NPE, PHENIX:NPE}. The paper is organized as follows. In Sec.~\ref{sec:experiment}, components of the STAR detector relevant to this analysis are briefly discussed. Section~\ref{sec:analysis} is dedicated to the details of the data analysis of inclusive $\rm HFE$ production. Finally, the results are reported and compared with published results and model calculations in Sec.~\ref{sec:results}.

\section{\label{sec:experiment}Experimental setup}
\subsection{Detector}

STAR ~\cite{ref:star_det} is a multi-purpose detector with a large acceptance at RHIC. In this analysis, three main STAR subsystems are used: the Time Projection Chamber (TPC)~\cite{ref:tpc_det}, the Barrel Electromagnetic Calorimeter (BEMC)~\cite{ref:bemc_det}, and the Beam-Beam Counters (BBC)~\cite{ref:bbc_det}. The TPC is a gas-filled detector providing tracking of charged particles with the pesudorapidity range of $|\eta| < 1$ and full azimuthal coverage. It is used for momentum determination and particle identification via energy loss measurement ($dE/dx$) for charged particles with $\pt > 0.2\, \rm GeV/c$. The BEMC is a lead-scintillator sampling calorimeter surrounding the TPC with a depth of 21 radiation lengths, covering the full azimuth ($\phi$) and $|\eta|< 1$. The BEMC is segmented into 4800 projective towers, each with a size of $0.05 \times 0.05$ in $\phi \times \eta$. It is used for electron identification and provides online triggers for high-$\pt$ electrons. The BBC, covering $3.3 <|\eta|< 5.0$, is located on both sides of the center of the detector at a distance of 3.75 m. Each BBC is made up of 18 hexagonal scintillator tiles. Signals in both BBCs form a prompt coincidence to provide a minimum bias trigger.
\subsection{Triggers and datasets}

The reported measurement of $\rm HFE$ production in $p$+$p$ collisions at \s{200} GeV utilizes data recorded by the STAR experiment in 2012 that satisfy the High Tower (HT) triggers in addition to the BBC minimum bias trigger condition. HT triggers require the transverse energy ($E_{\rm T}$) deposition in at least one single BEMC tower to pass a given Analog to Digital Converter (ADC) threshold. The tower ADC value is proportional to the $E_{\rm T}$ deposited by particles. Events used in this analysis are from two HT triggers with $E_{\rm T}$ thresholds of 2.6 GeV (HT0) and 4.2 GeV (HT2), which correspond to integrated luminosities of 1.4 and 23.5 pb$^{-1}$, respectively. In these HT triggered events, particle tracks in the TPC are projected onto the BEMC tower plane, and only those electron candidates whose projected trajectories can be associated to BEMC clusters that include trigger towers are selected. The results presented in this paper combine HT0 events for $\pt<$ 4.5 GeV/$c$ and HT2 for $\pt\geq$ 4.5 GeV/$c$.

\section{\label{sec:analysis}Data analysis for inclusive $\rm HFE$ production}

\subsection{Analysis principles}
Four steps are carried out to measure $\rm HFE$ production in this analysis: 
\begin{enumerate}
\item identification and purity correction of inclusive electrons ($\rm INE$),
\item identification and efficiency correction of the photonic electrons ($\rm PHE$), and subtraction of $\rm PHE$ from the $\rm INE$ sample,
\item efficiency correction of non-photonic electrons ($\rm NPE$),
\item subtraction of remaining background sources called hadron decayed electrons ($\rm HDE$), including di-electron decays of light vector mesons ($\rho$, $\omega$, $\phi$), quarkonium decays ($\jpsi$, $\Upsilon$),  Drell-Yan processes, and kaon semi-leptonic decays ($\Ke$).
\end{enumerate}
The first three steps can be summarized by\begin{equation}
N_{\rm NPE}= \frac{N_{\rm INE}\times P_{\rm e} -  N_{\rm PHE}/\effPho}{\epsilon_{\rm total}},
\label{eq:NPEyield}
\end{equation} 
where $N_{\rm NPE}$ is the non-photonic electron yield, $N_{\rm INE}$ is the inclusive electron candidate yield, $P_{\rm e}$ is the purity of the candidate electron sample, $N_{\rm PHE}$ is the photonic electron yield, $\effPho$ is the photonic electron identification efficiency, and $\epsilon_{\rm total}$ is the overall efficiency for triggering, tracking and particle identification of electrons.

In the first step, electron candidates are identified using combined information from the TPC and BEMC, and a purity correction to account for hadron contamination statistically (as described in Sec.~\ref{sec:purity}) is applied to obtain the inclusive electron sample ($N_{\rm INE}\times P_{\rm e}$). In the second step, the yield of $\rm PHE$, which is the main source of background in this analysis, is calculated. It consists of electrons from photon conversion in the detector material and Dalitz decays of $\pi^0$ and $\eta$ mesons ($\pi^0 \rightarrow \gamma e^+ e^-$, $\eta \rightarrow \gamma e^+ e^-$). The contribution of $\rm PHE$ is evaluated by reconstructing the di-electron mass ($M_{e^+e^-}$) spectrum (as described in Sec.~\ref{sec:pho:ele:id}). The observed PHE yield is corrected for PHE identification efficiency ($N_{\rm PHE}/\effPho$) and subtracted from the inclusive electron sample. In the third step, the remaining electrons in the inclusive electron sample are then corrected for single electron tracking, triggering, and identification efficiencies to obtain the $\rm NPE$ yield. In the last step, the $\rm HDE$ background  is subtracted, after which the electron sample is from open charm and bottom hadron decays.

\subsection{Event selection and electron identification}
During data processing, the event vertex is reconstructed offline in 3 dimensions, based on charged particle trajectories in the TPC, and it is called the primary vertex. In addition, at least two tracks contributing to vertex determination are required to either match to hits in the fast BEMC or Time-of-Flight~\cite{ref:tof} detectors or cross the TPC central membrane, to suppress pileup. To ensure a uniform TPC acceptance, only events with primary vertices located within $\pm$35~cm from the geometrical center of the TPC along the beam line direction and within 2~cm in the radial direction are selected.

A set of selection criteria is applied to ensure a high quality sample of tracks in the analysis. The number of points measured in the TPC (TPC hits) on a track is required to be at least 20 to ensure good tracking. At least 15 of these TPC hits must be used to measure the charged particle ionizing energy loss in the TPC gas to ensure a good $dE/dx$ resolution. The ratio of the number of used to the maximum possible number of TPC hits, which accounts, \textit{e.g.}, for inactive electronic channels, is required to be higher than 0.52 to avoid split tracks. The distance-of-closest-approach (DCA) between a track trajectory in the TPC and the primary vertex (gDCA) is required to be less than 1.5 cm to suppress background electrons produced in the detector material. Additional selections are applied to minimize photonic electron background from photon conversions in detector material. Tracks are required to have $|\eta|<0.7$ to avoid the beam pipe support structure. Also, to suppress photon conversion in the TPC gas, we require at least one hit within the first three TPC pad rows.

Electrons are identified using $dE/dx$ in the TPC and energy deposition in the BEMC. First, the BEMC clusters are associated with the TPC tracks by projecting track trajectories onto the BEMC tower plane, and electron candidates are required to have the momentum-to-energy ratio ($p/E$) from 0.3 to 1.5~\cite{ref:Zebo:thesis}, where energy is that of the most energetic tower in a BEMC cluster and momentum is measured by the TPC. Second, tracks with -0.5 $<\nse<$ 3 are selected, where $\nse$ represents a standardized energy loss expected for electrons. The $\nse$ is defined as $\nse = ln((dE/dx_{mea})/(dE/dx_{th}))/\sigma_{dE/dx}$, where $dE/dx_{mea}$ is the measured value, $dE/dx_{th}$ is the theoretical value for electrons and $\sigma_{dE/dx}$ is the experimental $dE/dx$ resolution. Electron candidates that pass all the aforementioned selection criteria are called the inclusive electrons. They are composed primarily of electrons, including $\rm HFE$, $\rm PHE$, and $\rm HDE$ sources, but also contain hadron contamination.

\subsection{\label{sec:purity}Electron purity}
\begin{figure}
\includegraphics[scale=0.3]{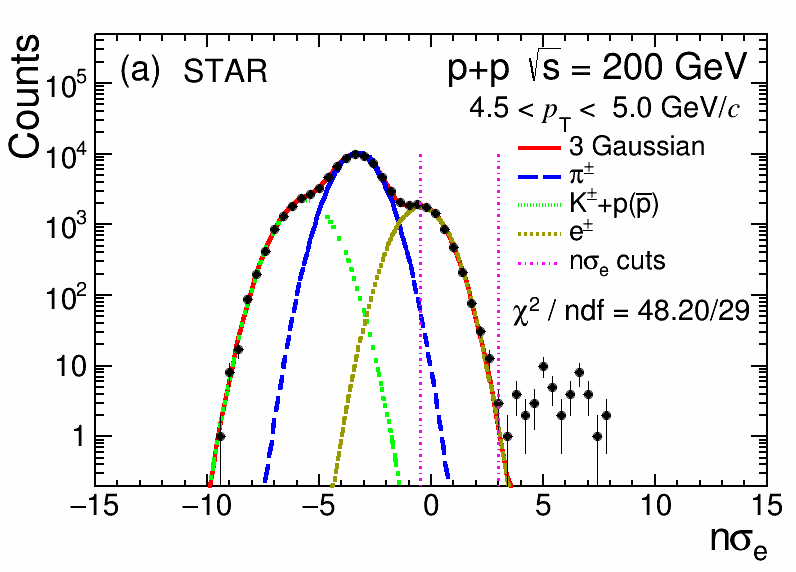}
\includegraphics[scale=0.3]{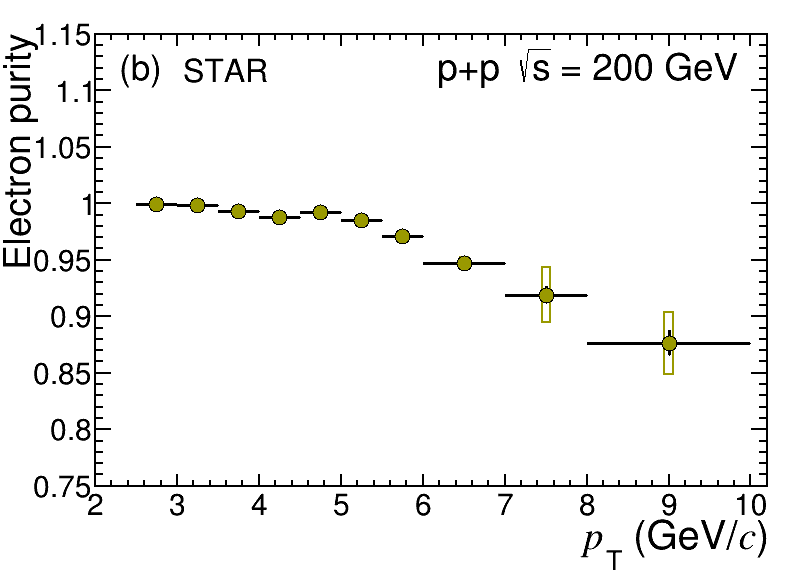}
\caption{\label{Fig:nseFits} (a) Example of $\nse$ distribution (black circles) with three-Gaussian fit (solid red curve) at 4.5 $<\pt<$ 5.0 GeV/$c$ in $p$+$p$ collisions at \s{200} GeV. Gaussian functions (dotted curves in various colors) represent fits for different particle species. The dotted pink vertical lines indicate the $-0.5< \nse <$  3 range used for electron selection. The small bump at 4 $< \nse <$ 10 is from the merging of two tracks~\cite{STAR:dielectronau}. (b) Electron purity as a function of $\pt$ in $p$+$p$ collisions at \s{200} GeV. The vertical bars represent statistical uncertainties while the boxes represent systematic uncertainties.} 
\end{figure}

The electron purity of the inclusive electron sample is estimated by a constrained fit to the $\nse$ distribution of inclusive electron candidates with the $p/E$ cut in each $\pt$ bin, prior to the $\nse$ selection being applied. Three Gaussian functions representing the distributions of $\pi^{\pm}$, $K^{\pm}$ + $p$($\bar{p}$), and $e^{\pm}$ are summed together to fit the $\nse$ distribution. The constraints on the Gaussian function representing electrons are obtained from the $\nse$ distribution of a pure electron sample, i.e. photonic electrons selected with a tight invariant mass cut $M_{e^+e^-}< 0.1\, \rm GeV/c^2$, as described in Sec.~\ref{sec:pho:ele:id}. For each $\pt$ bin, the pure electron $\nse$ distribution is fit with a single Gaussian function and the obtained mean and width are used to constrain the electron shape in the three-Gaussian fit. The $\nse$ distributions for hadrons are also expected to follow Gaussian distributions. The initial mean $\nse$ values are obtained from theoretical Bichsel function calculations~\cite{Bichsel:2006cs} and the initial widths are set to be one. Figure~\ref{Fig:nseFits} (a) shows an example of the three-Gaussian fit to the inclusive electron candidates at 4.5 $<\pt<$ 5.0 GeV/$c$. The purity is obtained by taking the ratio of the integral of the electron fit function to that of the overall fit function in the $\nse$ cut range (-0.5 $<\nse<$ 3). Figure~\ref{Fig:nseFits} (b) shows the purity of the inclusive electron sample as a function of $\pt$ with statistical and systematic uncertainties as described in Sec.~\ref{sec:results}.

\subsection{\label{sec:pho:ele:id}Photonic electron subtraction}

Photonic electrons arise from 2-body $\gamma$ conversions ($\gamma \rightarrow e^+e^-$) and 3-body Dalitz decays of $\pi^0$ and $\eta$ mesons ($\pi^0/ \eta \rightarrow e^+e^-\gamma$). Electrons among the inclusive electron sample, referred to as the tagged electrons in the following, are paired with oppositely charged tracks (partner electrons) in the TPC in the same event to reconstruct the invariant mass of their photonic parent, $\gamma$, $\pi^0$, or $\eta$. Such pairs are called unlike-sign (US) pairs. Unless specified otherwise, the $\pt$ assigned to an electron pair is that of its tagged electron. A looser set of quality cuts ($|\eta|< 1$, at least 15 TPC hits used for track reconstruction and $\pt> 0.3\, \rm GeV/c$), compared to the tagged electrons, is applied to select partner electrons, in order to enhance the probability of finding them. No $\nse$ or BEMC $p/E$ cuts are applied to the partner electrons since the invariant mass cut alone is sufficient to identify photonic electrons. One complication is that the primary momentum of a track, used for an electron candidate so far, is calculated with the assumption it originates from the primary vertex, and thus the primary vertex is included in the track trajectory. However, since photon conversions mostly take place in detector material away from the primary vertex, using the track momentum determined including the primary vertex will bias the reconstruction of photonic parents. Instead, the so-called global track momentum, which is calculated without including the primary vertex, is used. In order to ensure that a partner electron has the same origin as the tagged electron, a maximum DCA of 1.0 cm between two electron tracks is required.

\begin{figure}
\includegraphics[scale=0.3]{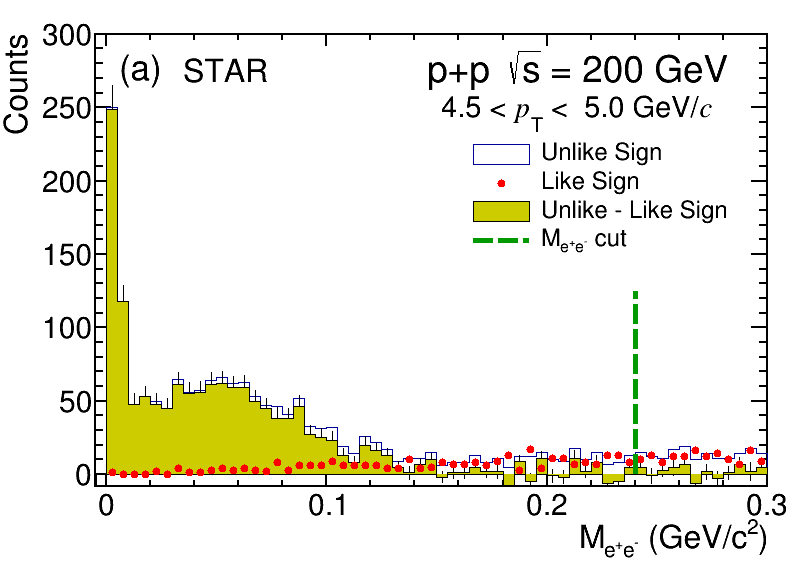}
\includegraphics[scale=0.3]{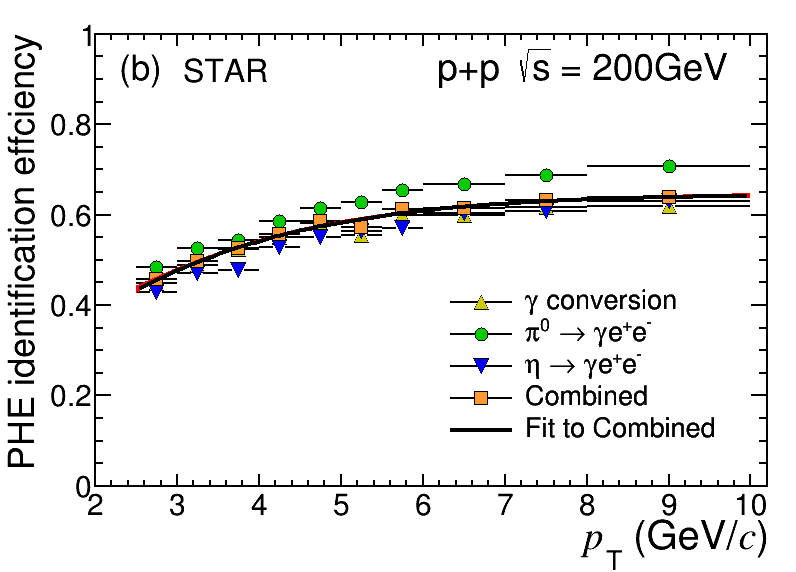}
\caption{\label{Fig:PhotoEleReco} (a) Example of invariant mass distribution for electron pairs at electron 4.5 $<\pt<$ 5.0 GeV/$c$ in $p$+$p$ collisions at \s{200} GeV. The blue histogram labeled ``Unlike Sign" shows the $e^+e^-$ pairs, the red circles labeled ``Like Sign" represent the combinatorial background, and the difference of these two is the photonic electrons, shown as the yellow histogram labeled ``Unlike-Like Sign". The dotted green vertical line indicates the photonic electron selection; (b) Combined photonic electron identification efficiency (orange squares) together with a fit (black curve) and parametrization uncertainty (red band narrower than the black curve), and individual photonic electron identification efficiencies: photon conversion (yellow up triangles), $\pi^{0} $ Dalitz decay (green circles), and $\eta$ Dalitz decay (blue down triangles) as a function of $\pt$ in $p$+$p$ collisions at \s{200} GeV.}
\end{figure} 

To account for the combinatorial background present in the selected $e^{+}e^{-}$ pairs, tagged electrons are also paired with same-charge partner electrons (like-sign (LS) pairs) in the same event. Figure~\ref{Fig:PhotoEleReco} (a) shows an example of invariant mass distributions for all $e^+e^-$ pairs and the combinatorial background at 4.5 $<\pt< 5.0\, \rm GeV/c$. The photonic electron yield is calculated as $\nPho = (N_{\rm US} - N_{\rm LS})$, where $N_{\rm US}$ and $N_{\rm LS}$ are the numbers of unlike-sign and like-sign tagged electrons with an invariant mass cut of $M_{e^+e^-}< 0.24\, \rm GeV/c^2$. Such a cut is chosen to account for the broadening of the invariant mass distribution at high tagged electron $\pt$. The invariant mass cut efficiency decreases from 99\% to 94\% with increasing $\pt$. 

The photonic electron identification efficiency, $\effPho$, which accounts for finding a partner electron and passing the pair DCA and invariant mass cuts, is calculated by propagating $\pi^0$, $\eta$ decays and $\gamma$ conversions through the GEANT [27] simulation of the STAR detector before embedding them into real events. The combined events then go through the same reconstruction and analysis software chain as the real data. Such events are called embedded events. The published $\eta$~\cite{ref:eta1,ref:eta2,ref:eta3,ref:eta4} and the average charged and neutral pion spectra~\cite{ref:eta1,ref:pi1,ref:pi2} are used as the inputs for $\eta$ and $\pi^0$ Dalitz decays. The input $\gamma$ $\pt$ spectrum is a sum of measured direct $\gamma$ by the PHENIX experiment~\cite{ref:dp1,ref:dp2,ref:dp3} and simulated $\pi^0 \rightarrow \gamma \gamma$/$e^+e^-\gamma$ and $\eta \rightarrow \gamma \gamma$/$e^+e^-\gamma$ processes using PYTHIA 6.419~\cite{ref:pythia} with default settings, in which the aforementioned $\pi^0$ and $\eta$ spectra are used as inputs. The rapidity distributions of $\pi^0$ and $\eta$ are parametrized with a Gaussian-like function $\cosh^{-2}\left( \frac{3y}{4\sigma(1-y^2/(2\sqrt{s}/m))}\right)$, where $\sigma=\sqrt{\ln(\sqrt{s}/(2m_{\rm N}))}$, $\sqrt{s}$ is a nucleon-nucleon center of mass energy, $m$ is the particle mass, and $m_{\rm N}$ is the nucleon mass~\cite{ref:INP, STAR:dielectron, CERES:dielectron}. All these photonic electron sources are then combined together, according to their yields. Figure~\ref{Fig:PhotoEleReco} (b) shows $\effPho$ as a function of $\pt$ for $\gamma$ conversion and two types of Dalitz decays in $p$+$p$ collisions. The $\pi^0$ Dalitz decays have higher efficiencies than other sources due to the narrower $e^+e^-$ invariant mass distributions and higher partner electron mean $\pt$. The combined $\effPho$, which starts from about 40\% at low $\pt$ and increases to about 60\% at $\pt \sim$ 10 GeV/$c$, is also shown in Fig.~\ref{Fig:PhotoEleReco} (b), along with a fit using the functional form $A/(e^{-(p_{\rm T}-p_{\rm 0})/p_{\rm 1}}+1)+C$, where $A$, $p_{\rm 0}$, $p_{\rm 1}$, and $C$ are free parameters.

\begin{figure}
\includegraphics[scale=0.3]{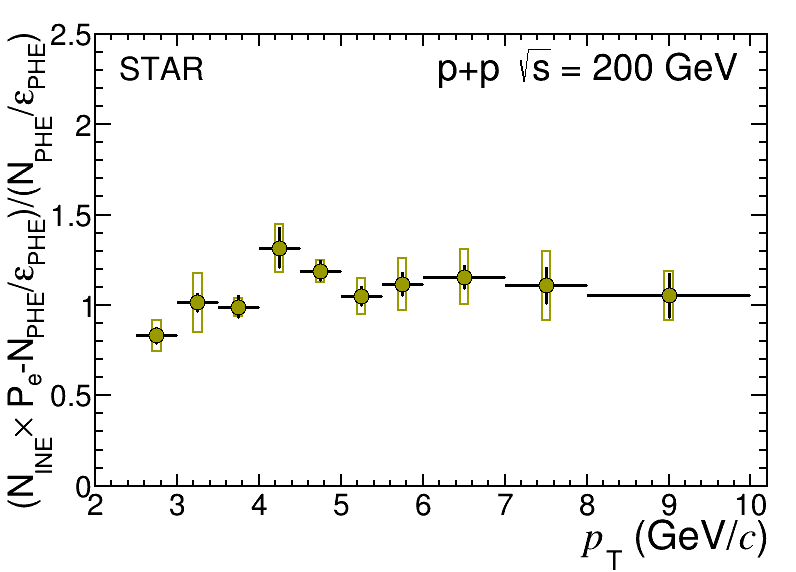}
\caption{\label{Fig:Purity_StoB_ratio} Signal-to-background ratio as a function of $\pt$, where the signals are non-photonic electrons [$N_{\rm INE}\times P_{\rm e} -  N_{\rm PHE}/\effPho$ in Eq. (\ref{eq:NPEyield})] and the backgrounds are photonic electrons ($N_{\rm PHE}/\effPho$ in Eq. [\ref{eq:NPEyield})], in $p$+$p$ collisions at \s{200} GeV. The vertical bars represent statistical uncertainties while the boxes represent systematic uncertainties (details in Sec.~\ref{sec:results}).} 
\end{figure}

After statistical subtraction of hadron contamination and photonic electrons, the remaining electrons are the non-photonic ones. Figure~\ref{Fig:Purity_StoB_ratio} shows the yield ratio of non-photonic electrons to photonic electron background as a function of $\pt$.

\begin{figure}
\includegraphics[scale=0.3]{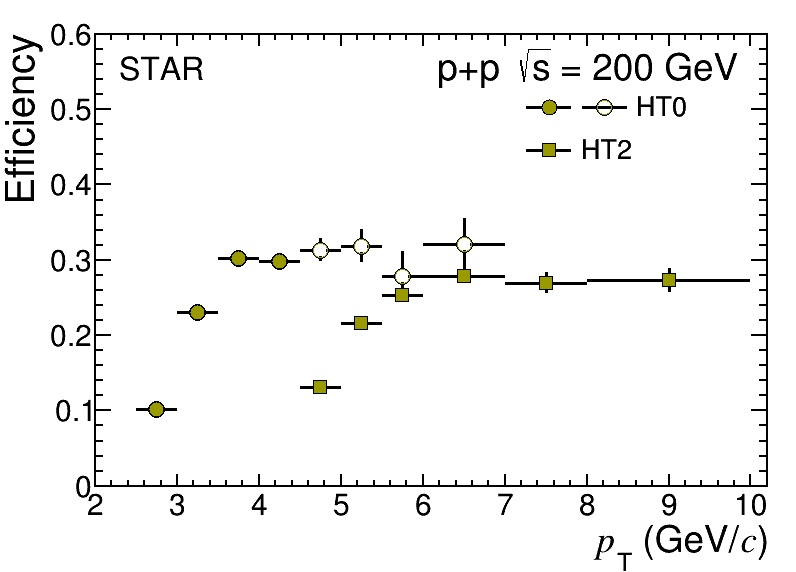} 
\caption{\label{Fig:Efficiency} Overall electron detection efficiency ($\epsilon_{\rm total}$ in Eq. [\ref{eq:NPEyield})] as a function of $\pt$ in $p$+$p$ collisions at \s{200} GeV. The circles and squares are the efficiencies for HT0- and HT2-triggered electrons, respectively. The vertical bars represent uncertainties. The solid points are used to correct the non-photonic electron yield.} 
\end{figure}

\subsection{\label{sec:eff}Track reconstruction and electron identification efficiency}

Before removing the $\rm HDE$ background, the non-photonic electron sample needs to be corrected for the overall efficiency [$\epsilon_{\rm total}$ in Eq. (\ref{eq:NPEyield})] of triggering, tracking reconstruction and electron identification. This overall efficiency is studied with a combination of data-driven approaches and utilization of the embedded events of single electrons. The combined detector acceptance and tracking efficiency is studied based on the embedded events of single electrons, which is about 69\% at low $\pt$ and 70\% at high $\pt$. The $\nse$ cut efficiency is calculated based on the Gaussian fit to the pure electron sample, as described in Sec.~\ref{sec:purity}, which decreases with $\pt$ and is about 55\%-50\%. The BEMC particle identification (PID) and trigger efficiencies are studied in embedded events with GEANT simulated BEMC responses to electrons. The BEMC PID efficiency is evaluated by taking the ratio of electrons with and without the BEMC selection, and is roughly 79\%. The trigger efficiency is obtained by requiring the offline ADC value of the most energetic tower in a BEMC cluster, matched to an electron track, to be larger than the threshold, and about 35\% at low $\pt$ and increases to about 99\% at $\pt\sim$ 10 GeV/c. Figure~\ref{Fig:Efficiency} shows the overall efficiency as a function of $\pt$ for HT0- and HT2-triggered electrons. It is about 10$-$30\% and increases with $\pt$. The solid points are used in this analysis, while the open points are only for comparison purpose.

\subsection{\label{sec:hde}Hadron decayed electron background}

Electrons from di-electron decays of light vector mesons ($\rho$, $\omega$, $\phi$), heavy quarkonium decays ($\jpsi$, $\Upsilon$), Drell-Yan processes, and kaon semi-leptonic decays ($\Ke$) are additional sources of background which need to be subtracted in order to obtain electrons from semileptonic decays of open heavy flavors. 

\begin{figure}
\includegraphics[scale=0.3]{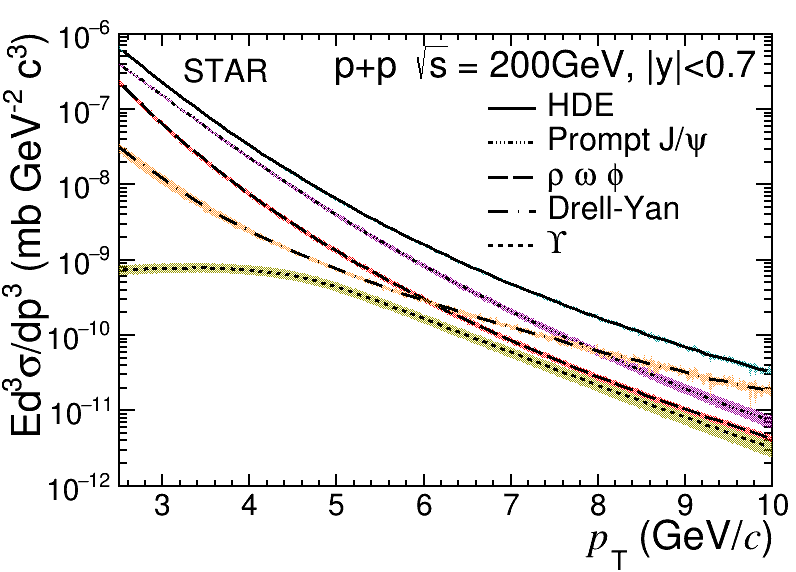}
\caption{\label{Fig:HDE} Invariant cross sections of the electrons from decays of prompt $\jpsi$ (dot-dashed line), $\Upsilon$ (dotted line), Drell-Yan (long dash-dotted line), light vector mesons (long dashed line) and the combined HDE contributions (solid line) in $p$+$p$ collisions at \s{200} GeV. The bands represent systematic uncertainties (details in Sec.~\ref{sec:results}).} 
\end{figure}

Inclusive $\jpsi$ spectra have been measured in 200 GeV $p$+$p$ collisions at mid-rapidity by both the STAR~\cite{STAR:jpsi} and PHENIX~\cite{PHENIX:jpsi} collaborations, and the combined $\jpsi$ $\pt$ spectra are parametrized with the Tsallis statistics~\cite{ref:TS, ref:TS1, ref:TS2}. The $\jpsi$ rapidity distribution is from PYTHIA~\cite{STAR:jpsip}. Since decayed electrons from non-prompt $\jpsi$ are one of the components of bottom-decayed electrons, the FONLL+CEM calculations~\cite{ref:fcem, ref:QCDCB} are used to remove the non-prompt $\jpsi$ contribution from the inclusive $\jpsi$ spectrum. The fraction of non-prompt $\jpsi$ to inclusive $\jpsi$ starts from 0.02 ($\pm 0.01$) and increases with increasing $\jpsi$ $\pt$, becoming up to 0.18 ($\pm 0.06$) for $\pt=11.75\, \rm{GeV}/c$. $\jpsi$ mesons are generated according to the parametrized $\pt$ spectrum, after removing the non-prompt $\jpsi$ contribution, and EvtGen~\cite{ref:evtgen} is utilized to describe their decays to electrons. In this procedure, $\jpsi$ is assumed to be unpolarized, which is consistent with the STAR measurement~\cite{STAR:jpsipnew}. The $\jpsi$ decayed electron cross section is represented by the dot-dashed line in Fig.~\ref{Fig:HDE}.

The $\Upsilon$ decayed electron contribution is estimated in a similar way as that for the $\jpsi$ except that the $\Upsilon$ spectrum and rapidity distributions are used as inputs. The $\pt$ spectra of $\Upsilon$ states are parametrized with the following function: $f=C \times \frac{\pt}{e^{\frac{\pt}{T}}+1}$, where the values of the $T$ and $C$ (free parameters) are taken from Ref.~\cite{ref:Pengfei:thesis}. The rapidity distribution of $\Upsilon$ is parametrized with the Gaussian-like function mentioned in Sec.~\ref{sec:pho:ele:id}. The cross section of the electrons from $\Upsilon$ decay is represented by the dotted line in Fig.~\ref{Fig:HDE}. 

The vector meson spectra are obtained through $m_{\rm T}$ ($\sqrt{\pt^2+m^2}$) scaling of the $\pi^0$ $\pt$-shape, i.e. replacing the $\pt$ with $\sqrt{\pt^2+m_{\rm m}^2-m_{\rm \pi^0}^2}$ in the fit function to the $\pi^0$  spectrum, where $m_{\rm m}$ is the mass of the vector meson. The absolute yield is determined by matching the ratio of vector meson over $\pi^0$ to the measured values at high $\pt$~\cite{STAR:dielectronp,PHENIX:NPE}. Their rapidity distributions are also obtained from calculation of the Gaussian-like function mentioned in Sec.~\ref{sec:pho:ele:id}. EvtGen is used to decay $\omega$ and $\phi$, while PYTHIA 6.419 with default settings is used to decay $\rho$ since EvtGen doesn't provide the electron decay channel for $\rho$. The following decay channels $\rho \rightarrow e^{+}e^{-}$, $\omega \rightarrow e^{+}e^{-}$, $\omega \rightarrow \pi^{0}e^{+}e^{-}$, $\phi \rightarrow e^{+}e^{-}$, and $\phi \rightarrow \eta e^{+}e^{-}$ are included in the calculation, and the resulting decayed electron cross section is shown as the long dashed line in Fig.~\ref{Fig:HDE}.

The Drell-Yan contribution is estimated by the PYTHIA simulation, which has the same settings as those in the PHENIX Drell-Yan measurement~\cite{PHENIX:drellyan}, and is shown as the long dash-dotted line in Fig.~\ref{Fig:HDE}. Furthermore, STAR simulation studies find that the $\Ke$ contribution is less than 1\% at $\pt>2$~GeV/c in $p$+$p$ collisions at \s{200} GeV~\cite{ref:Wenqin:thesis} and thus neglected. The overall $\rm HDE$ contribution, represented by the solid line in Fig.~\ref{Fig:HDE}, is subtracted from the NPE sample. This amounts to a ∼16\% reduction to the NPE yield integrated over the measured $\pt$ region. The remaining electrons are $\rm HFE$ reported in Sec.~\ref{sec:results}. The uncertainty of the $\rm HDE$ contribution will be discussed in Sec.~\ref{sec:results}. 

\section{\label{sec:results}Results}

The $\rm HFE$ cross section in \pp\ collisions is obtained as:
 \begin{equation}
{E\frac{d^3\sigma}{dp^3}(\rm HFE)}=\frac{1}{2}\frac{1}{L}\frac{N_{\rm NPE}}{2\pi \pt \Delta \pt \Delta y}-E\frac{d^3\sigma}{dp^3}(\rm HDE),
\end{equation}
where $L$ is the integrated luminosity, $\pt$ is the weighted average of the bin, $\Delta \pt$ and $\Delta y$ are the $\pt$ and rapidity intervals, respectively. $L=\frac{N_{\rm events}}{\sigma_{\rm NSD}}$, where $N_{\rm events}$ is the equivalent number of minimum bias events of the triggered data and $\sigma_{\rm NSD}$ is the non-singly diffractive cross section ($\sigma_{\rm NSD}=30.0 \pm 2.4$ mb~\cite{STAR:D0}). 

The total systematic uncertainty is obtained as the square root of the quadratic sum of the individual systematic uncertainties discussed below. The uncertainties in the $\rm NPE$ reconstruction efficiency are estimated by changing the following selections in data and simulation simultaneously: (i) the number of TPC hits used for track reconstruction and $dE/dx$ calculation from 20 and 15 to 25 and 18 (the larger impact of the two variations on the NPE yield is used); (ii) gDCA from $1.5\, \rm cm$ to $1.0\, \rm cm$; and (iii) $0.3 <p/E< 1.5$ to 0.6 $<p/E<$ 1.5 or $0.3 <p/E< 1.8$, as well as by changing only in the simulation ADC thresholds for HT triggers by $\pm$ 3.5\%. The uncertainty on electron purity is estimated based on the mean and width uncertainties of the Gaussian fit to the pure electron $\nse$ distribution, in which the mean and width are varied independently within one sigma. The PHE identification efficiency uncertainty stems from the uncertainties in the parametrization of $\rm PHE$ identification efficiency, parametrizations of $\pi^0$ and $\eta$ spectra, branching ratios of electrons from $\pi^0$ and $\eta$ decays, tracking efficiency of partner electrons and variations in the PHE selection criteria, \textit{i.e.}, changing  $M_{e^+e^-}<0.24\, \rm GeV/c^2$ to $M_{e^+e^-}< 0.15\, \rm GeV/c^2$ and partner electron $\pt$ from 0.3 GeV/$c$ to 0.2 GeV/$c$. parametrization uncertainties are taken as the 68\% confidence interval of the fit functions. Such an approach is also used in estimating the parametrization uncertainties described in the following. The uncertainty of the $\nse$ cut efficiency is estimated from the parameter errors in fitting the pure electron $\nse$ distribution with a Gaussian function, taking into account the correlation between the mean and the width, and from varying the INE selection cut from $-0.5<\nse<3.0$ to $0.0<\nse<3.0$. The uncertainty of the $\rm HDE$ contribution includes those from $\jpsi$, $\Upsilon$, vector meson, and Drell-Yan contributions. The parametrization uncertainty for the inclusive $\jpsi$ spectrum and the uncertainty from FONLL+CEM calculations of the non-prompt $\jpsi$ contribution are taken into account. So are the uncertainties in the $\Upsilon$ yield and spectrum shape~\cite{ref:Pengfei:thesis}. The uncertainties in the parametrization of the $\pi^0$ spectrum as well as in the measured yield ratios of vector mesons to $\pi^0$ are also propagated to the decayed electron cross section. The uncertainty from the Drell-Yan contribution is estimated using the same method as in the PHENIX published Drell-Yan result~\cite{PHENIX:drellyan}. The uncertainty in the BBC trigger and vertex reconstruction efficiencies, amounting to 4.9-5.2\%, arises from the event multiplicity difference in data and simulation, the difference in the versions used (PYTHIA 6 vs PYTHIA 8), and the different parameter settings in the simulation (PYTHIA 8.1.62~\cite{ref:pythia8} default setting with the STAR heavy flavor tune~\cite{ref:jpsipppau} vs PYTHIA 8.1.62 4CX~\cite{ref:4cx} setting with the STAR heavy flavor tune). The global uncertainty from the luminosity determination is 8\%~\cite{STAR:D0}. Table \ref{tab:syserror} summarizes the size of the uncertainties from the different sources and the total uncertainty. 
\begin{table}
\caption{\label{tab:syserror}Summary of systematic uncertainties, in percentage, for the $\rm HFE$ cross section. A range is given if the uncertainty varies with $\rm HFE$ $\pt$.
}
\begin{ruledtabular}
\begin{tabular}{p{175pt}<{\centering}p{65pt}<{\centering}}
\textrm{Source}&
\textrm{Uncertainty}\\
\colrule
$\rm NPE$ reconstruction efficiency & 1.6-7.8\%\\
Electron purity extraction & 0.1-7.3\%\\
$\rm PHE$ identification efficiency & 2.2-7.3\%\\
$\nse$ cut efficiency& 2.5-10.8\%\\
BBC trigger and vertex reconstruction efficiencies & 4.9-5.2\% \\
$\rm HDE$ contribution & 0.7-1.4\%\\
Luminosity & 8\%\\
Total & 10.4\%-17.4\%\\
\end{tabular}
\end{ruledtabular}
\end{table}

\begin{figure}
\includegraphics[scale=0.3]{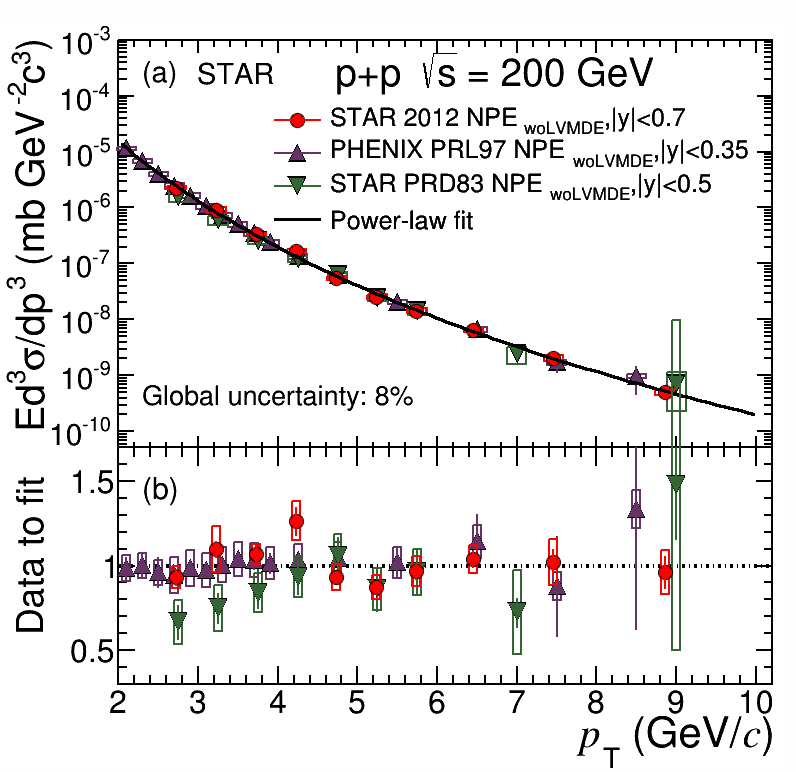}
\caption{\label{Fig:NPE} (a) The $\rm NPE$ cross section after subtracting the light vector meson contribution at STAR in $p$+$p$ collisions at \s{200} GeV from 2012 (filled circles) along with published STAR data from 2005 and 2008 (filled down triangles)~\cite{STAR:NPE}, published PHENIX data from 2005 (filled up triangles)~\cite{PHENIX:NPE} and a power-law fit (curve). (b) Ratio of data over power-law fit. The vertical bars and the boxes represent statistical and systematic uncertainties, respectively.}
\end{figure}

\begin{figure}[htbp]
\includegraphics[scale=0.3021]{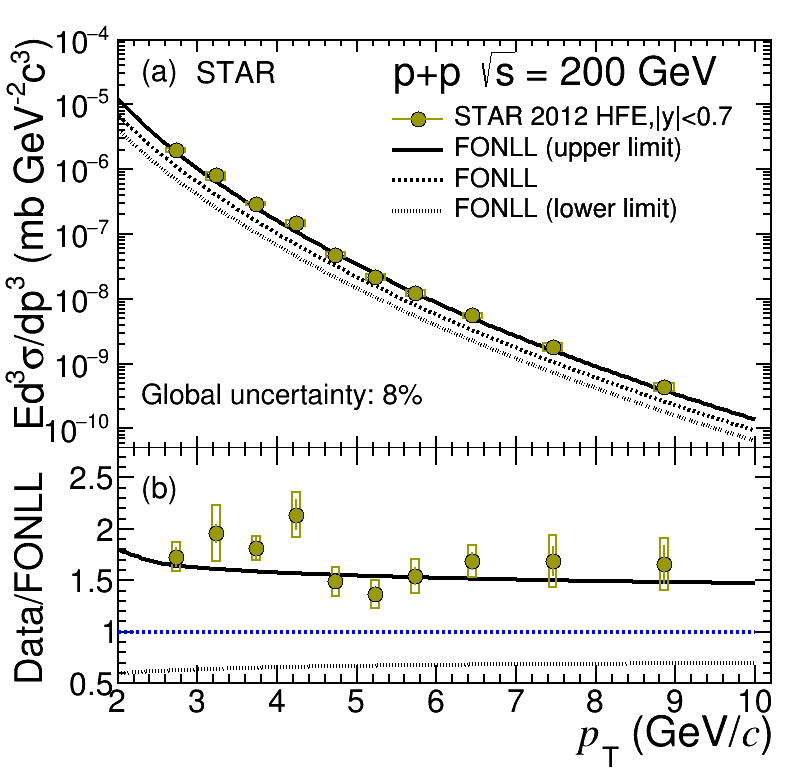}
\caption{\label{Fig:HFE} (a) The $\rm HFE$ cross section at STAR in $p$+$p$ collisions at \s{200} GeV from 2012 (filled circles) and the FONLL calculation (curves). (b) Ratio of data over FONLL calculation. The vertical bars and the boxes represent statistical and systematic uncertainties, respectively.}
\end{figure}

In order to compare with the published STAR~\cite{STAR:NPE} and PHENIX~\cite{PHENIX:NPE} results, where electrons from heavy quarkonium decays and Drell-Yan process were not subtracted, only electrons from the light vector meson ($\rho$, $\omega$, $\phi$) decays are subtracted from NPE:
 \begin{equation}
{E\frac{d^3\sigma}{dp^3}(\rm NPE_{\rm woLVMDE})}=\frac{1}{2}\frac{1}{L}\frac{N_{\rm NPE}}{2\pi \pt \Delta \pt \Delta y}-E\frac{d^3\sigma}{dp^3}(\rm LVMDE),
\end{equation}
where $E\frac{d^3\sigma}{dp^3}(\rm LVMDE)$ is the cross section of electrons from light vector meson decays.
The result is shown in Fig.~\ref{Fig:NPE} (a), together with the previously published results. A combined power-law ($f(\pt) =A/(1+\pt/B)^n$, where $A$, $B$ and $n$ are free parameters) fit to the PHENIX data and the result presented in this paper gives the power $n=8.99 \pm 0.26$, and is shown in Fig.~\ref{Fig:NPE} (a). Ratios of different results to the power-law fit are plotted in Fig.~\ref{Fig:NPE} (b). Overall, there is a good agreement among these results within their uncertainties. The new result is measured with significantly improved precision relative to the previous measurements at $\pt >6\, \rm GeV/c$. Figure~\ref{Fig:HFE} (a) shows the measured $\rm HFE$ cross section for $p$+$p$ collisions at \s{200} GeV, compared with the FONLL calculation. The ratio of $\rm HFE$ data to the FONLL calculation is shown in Fig.~\ref{Fig:HFE} (b). The result reported in this paper is consistent with the FONLL prediction within uncertainties, but the central values sit at the upper limit of the theory uncertainty.

\section{\label{sec:summary}Summary}
The measurement of the cross section for production of electrons from open-charm and open-bottom hadron decays for 2.5 $<\pt<$ 10 GeV/$c$ in $p$+$p$ collisions at \s{200} GeV is reported. The result without subtracting the $\jpsi$, $\Upsilon$, and Drell-Yan contributions is consistent with the STAR and PHENIX published results, and significantly improved precision relative to the previous measurements is seen above 6 GeV/c. The result, with all background hadronic decay sources removed, is qualitatively consistent with the FONLL upper limit and provides further constraints on theoretical calculations. Furthermore, this result provides a precise reference for nuclear modification factor measurements for heavy flavor decayed electrons in heavy-ion collisions. It also facilitates a study on the separation of the charm and bottom contributions in HFE in $p$+$p$ collisions~\cite{ref:btoe}.

\section{\label{sec:acknowledgement}Acknowledgment}
We thank the RHIC Operations Group and RCF at BNL, the NERSC Center at LBNL, and the Open Science Grid consortium for providing resources and support. This work was supported in part by the Office of Nuclear Physics within the U.S. DOE Office of Science, the U.S. National Science Foundation, the Ministry of Education and Science of the Russian Federation, National Natural Science Foundation of China, Chinese Academy of Science, the Ministry of Science and Technology of China and the Chinese Ministry of Education, the Higher Education Sprout Project by Ministry of Education at NCKU, the National Research Foundation of Korea, Czech Science Foundation and Ministry of Education, Youth and Sports of the Czech Republic, Hungarian National Research, Development and Innovation Office, New National Excellency Programme of the Hungarian Ministry of Human Capacities, Department of Atomic Energy and Department of Science and Technology of the Government of India, the National Science Centre of Poland, the Ministry  of Science, Education and Sports of the Republic of Croatia, RosAtom of Russia and German Bundesministerium f\"ur Bildung, Wissenschaft, Forschung and Technologie (BMBF), Helmholtz Association, Ministry of Education, Culture, Sports, Science, and Technology (MEXT) and Japan Society for the Promotion of Science (JSPS).

\nocite{*}

\bibliography{main}

\end{document}

%% file: star-author-list-2021-07-14.aps.tex
\affiliation{Abilene Christian University, Abilene, Texas   79699}
\affiliation{AGH University of Science and Technology, FPACS, Cracow 30-059, Poland}
\affiliation{Alikhanov Institute for Theoretical and Experimental Physics NRC "Kurchatov Institute", Moscow 117218, Russia}
\affiliation{Argonne National Laboratory, Argonne, Illinois 60439}
\affiliation{American University of Cairo, New Cairo 11835, New Cairo, Egypt}
\affiliation{Brookhaven National Laboratory, Upton, New York 11973}
\affiliation{University of California, Berkeley, California 94720}
\affiliation{University of California, Davis, California 95616}
\affiliation{University of California, Los Angeles, California 90095}
\affiliation{University of California, Riverside, California 92521}
\affiliation{Central China Normal University, Wuhan, Hubei 430079 }
\affiliation{University of Illinois at Chicago, Chicago, Illinois 60607}
\affiliation{Creighton University, Omaha, Nebraska 68178}
\affiliation{Czech Technical University in Prague, FNSPE, Prague 115 19, Czech Republic}
\affiliation{Technische Universit\"at Darmstadt, Darmstadt 64289, Germany}
\affiliation{ELTE E\"otv\"os Lor\'and University, Budapest, Hungary H-1117}
\affiliation{Frankfurt Institute for Advanced Studies FIAS, Frankfurt 60438, Germany}
\affiliation{Fudan University, Shanghai, 200433 }
\affiliation{University of Heidelberg, Heidelberg 69120, Germany }
\affiliation{University of Houston, Houston, Texas 77204}
\affiliation{Huzhou University, Huzhou, Zhejiang  313000}
\affiliation{Indian Institute of Science Education and Research (IISER), Berhampur 760010 , India}
\affiliation{Indian Institute of Science Education and Research (IISER) Tirupati, Tirupati 517507, India}
\affiliation{Indian Institute Technology, Patna, Bihar 801106, India}
\affiliation{Indiana University, Bloomington, Indiana 47408}
\affiliation{Institute of Modern Physics, Chinese Academy of Sciences, Lanzhou, Gansu 730000 }
\affiliation{University of Jammu, Jammu 180001, India}
\affiliation{Joint Institute for Nuclear Research, Dubna 141 980, Russia}
\affiliation{Kent State University, Kent, Ohio 44242}
\affiliation{University of Kentucky, Lexington, Kentucky 40506-0055}
\affiliation{Lawrence Berkeley National Laboratory, Berkeley, California 94720}
\affiliation{Lehigh University, Bethlehem, Pennsylvania 18015}
\affiliation{Max-Planck-Institut f\"ur Physik, Munich 80805, Germany}
\affiliation{Michigan State University, East Lansing, Michigan 48824}
\affiliation{National Research Nuclear University MEPhI, Moscow 115409, Russia}
\affiliation{National Institute of Science Education and Research, HBNI, Jatni 752050, India}
\affiliation{National Cheng Kung University, Tainan 70101 }
\affiliation{Nuclear Physics Institute of the CAS, Rez 250 68, Czech Republic}
\affiliation{Ohio State University, Columbus, Ohio 43210}
\affiliation{Institute of Nuclear Physics PAN, Cracow 31-342, Poland}
\affiliation{Panjab University, Chandigarh 160014, India}
\affiliation{Pennsylvania State University, University Park, Pennsylvania 16802}
\affiliation{NRC "Kurchatov Institute", Institute of High Energy Physics, Protvino 142281, Russia}
\affiliation{Purdue University, West Lafayette, Indiana 47907}
\affiliation{Rice University, Houston, Texas 77251}
\affiliation{Rutgers University, Piscataway, New Jersey 08854}
\affiliation{Universidade de S\~ao Paulo, S\~ao Paulo, Brazil 05314-970}
\affiliation{University of Science and Technology of China, Hefei, Anhui 230026}
\affiliation{Shandong University, Qingdao, Shandong 266237}
\affiliation{Shanghai Institute of Applied Physics, Chinese Academy of Sciences, Shanghai 201800}
\affiliation{Southern Connecticut State University, New Haven, Connecticut 06515}
\affiliation{State University of New York, Stony Brook, New York 11794}
\affiliation{Instituto de Alta Investigaci\'on, Universidad de Tarapac\'a, Arica 1000000, Chile}
\affiliation{Temple University, Philadelphia, Pennsylvania 19122}
\affiliation{Texas A\&M University, College Station, Texas 77843}
\affiliation{University of Texas, Austin, Texas 78712}
\affiliation{Tsinghua University, Beijing 100084}
\affiliation{University of Tsukuba, Tsukuba, Ibaraki 305-8571, Japan}
\affiliation{Valparaiso University, Valparaiso, Indiana 46383}
\affiliation{Variable Energy Cyclotron Centre, Kolkata 700064, India}
\affiliation{Warsaw University of Technology, Warsaw 00-661, Poland}
\affiliation{Wayne State University, Detroit, Michigan 48201}
\affiliation{Yale University, New Haven, Connecticut 06520}

\author{M.~S.~Abdallah}\affiliation{American University of Cairo, New Cairo 11835, New Cairo, Egypt}
\author{B.~E.~Aboona}\affiliation{Texas A\&M University, College Station, Texas 77843}
\author{J.~Adam}\affiliation{Brookhaven National Laboratory, Upton, New York 11973}
\author{L.~Adamczyk}\affiliation{AGH University of Science and Technology, FPACS, Cracow 30-059, Poland}
\author{J.~R.~Adams}\affiliation{Ohio State University, Columbus, Ohio 43210}
\author{J.~K.~Adkins}\affiliation{University of Kentucky, Lexington, Kentucky 40506-0055}
\author{G.~Agakishiev}\affiliation{Joint Institute for Nuclear Research, Dubna 141 980, Russia}
\author{I.~Aggarwal}\affiliation{Panjab University, Chandigarh 160014, India}
\author{M.~M.~Aggarwal}\affiliation{Panjab University, Chandigarh 160014, India}
\author{Z.~Ahammed}\affiliation{Variable Energy Cyclotron Centre, Kolkata 700064, India}
% ...

\collaboration{STAR Collaboration}\noaffiliation